\documentclass[nofootinbib,superscriptaddress,twocolumn,aps,prl,showpacs,10pt]{revtex4-1}

\usepackage{amsmath}
\usepackage{amssymb}
\usepackage[dvipdfmx]{graphicx}
\usepackage{mathrsfs}
\usepackage{color}

\begin{document}
\title{Microscopic phase-space exploration modeling of $^{258}$Fm spontaneous fission }

\author{Yusuke Tanimura} \email{tanimura@ipno.in2p3.fr}
\affiliation{Institut de Physique Nucl\'eaire, IN2P3-CNRS, Universit\'e Paris-Sud, Universit\'e Paris-Saclay, F-91406 Orsay Cedex, France}
\author{Denis Lacroix} \email{lacroix@ipno.in2p3.fr}
\affiliation{Institut de Physique Nucl\'eaire, IN2P3-CNRS, Universit\'e Paris-Sud, Universit\'e Paris-Saclay, F-91406 Orsay Cedex, France}
\author{Sakir Ayik}
\affiliation{Physics Department, Tennessee Technological University, Cookeville, TN 38505, USA}

\date{\today}

\begin{abstract}
We show that the total kinetic energy (TKE) of nuclei after the spontaneous fission of $^{258}$Fm can be well reproduced using 
simple assumptions on the quantum collective phase-space explored by the nucleus after passing the fission barrier.  Assuming 
energy conservation and phase-space exploration according to the stochastic mean-field approach, a set of initial densities is generated.
Each density is then evolved in time using the nuclear time-dependent density-functional theory with pairing. This approach 
goes beyond mean-field by allowing spontaneous symmetry breaking as well as a wider dynamical phase-space exploration
leading to larger fluctuations in collective space. The total kinetic energy and mass distributions are calculated. New information on the fission process: fluctuations in scission time, strong correlation between TKE and collective deformation
as well as pre-scission particle emission, are obtained. 
We conclude that fluctuations of TKE and mass are triggered by quantum fluctuations.
\end{abstract}


\maketitle

The dynamical modeling of a nuclear Fermi quantum droplet that spontaneously breaks into two pieces 
represents one of the most exciting challenges of nuclear physics today. Beside its description, a deeper 
microscopic understanding of spontaneous fission (SF) is of great importance to form the heaviest elements at 
the frontier of the nuclear chart \cite{Sea90,Oga07,Sob07} or to further improve our knowledge about the competition between the r-process and fission
during the primordial nucleosynthesis \cite{Rol88}. Motivated also
by its importance in nuclear energy production, intensive experimental 
efforts have been made to accumulate precise measurements \cite{Sch00,Sch01,Thi02,Esc12,And13}.
In recent years, an increasing effort was made to remove empirical ingredients that are 
employed in macroscopic modeling of fission and use microscopic theories \cite{Kra12,Rob15}.  The nuclear Density Functional Theory
(DFT) 
is a suitable starting point to describe some aspects related to 
the large amplitude collective motion (LACM). A minimal information obtained from DFT is the adiabatic
collective energy landscape \cite{Rob15}. One challenge to describe spontaneous fission (SF)
is the necessity to explicitly treat the evolution as a {\it quantum} dynamic in collective space. Progresses 
have been made with the Time-Dependent Generator Coordinate Method (TDGCM) \cite{Rei87,Gou05,Reg16}. This theory by treating quantally collective  degrees of freedom (DOF) is promising. However, 
the adiabatic assumption often made becomes critical especially close to scission \cite{Tan15}. Dissipation of the 
collective motion into internal excitations also plays a key role to understand the excitation energy and kinetic energy 
sharing during fragments separation \cite{Abe96}. Understanding this dissipation requires to 
include many-body states beyond the adiabatic limit \cite{Ber11}. It is yet unclear how the pre-scission neutron and proton emissions 
can be  incorporated  in the TDGCM. 

The nuclear time-dependent DFT (TDDFT) overcomes some of these limitations. With recently developed symmetry unrestricted codes, LACM with arbitrary shapes \cite{Kim97,Sim01,Nak05,Mar05,Uma05,Was08,Ste11,Sca13,Gao14} can be described including 
one-body dissipation as well as particle
evaporation. As noted in \cite{Sim14}, it  can describe the average kinetic energy of fragments after fission. The inclusion of pairing effects significantly extend the applicability of this approach \cite{Sca15, Tan15}. As shown in Ref. \cite{Bul16} using TDHFB, the treatment of dynamical pairing has solved the threshold anomaly \cite{Sca15,Tan15, God15,God16},  i.e. the fact that in a range of 
deformation larger than the barrier position, 
heavy systems were not fissioning in TDDFT when pairing was neglected.
Contrary to our earlier 
belief  \cite{Sca15}, this problem is also solved using the TDHF+BCS approximation.
TDDFT with pairing has also its intrinsic limitation that prevents it to properly describe the  fission: (i) although it is a quantum theory in single-particle space, it gives a quasi-classical description of the collective motion. 
As a consequence, fluctuations of one-body DOFs are strongly underestimated. (ii) The absence of spontaneous symmetry breaking also prevents proper description of fission \cite{Neg89}. 

We propose a novel method able to describe quantum fluctuations and spontaneous symmetry breaking 
together with the possibility to obtain fully microscopically fragment mass  and TKE distributions
in SF.  The method we use to go beyond mean-field theory is based on the fact that quantum and thermal effects can be simulated 
by a sampling of initial conditions followed by quasi-classical evolutions, here TDDFT being considered as such. Similar strategy were used with success in quantum optics \cite{Gar04}, cold atoms \cite{Sin02} or more recently particle physics \cite{Gel16}. In nuclear physics this approach, called stochastic mean field \cite{Ayi08}, 
has been originally introduced such that initial fluctuations in collective space are reproduced
through fluctuations of the initial one-body density. In a series of works, it was shown that it surprisingly well accounts for correlations beyond mean-field \cite{Lac12,Lac13,Lac14b} while treating properly the dynamic close to a spontaneous symmetry breaking \cite{Lac12}. 
It is also able to include approximately many-body correlation to all orders by connecting the evolution to the BBGKY hierarchy \cite{Lac15}. 
Some formal and practical aspects are reviewed in Ref. \cite{Lac14}. Although the stochastic mean-field technique was applied to simple models \cite{Lac12,Lac13, Lac14b,Yil14} or to obtain expressions of transport coefficients  \cite{Ayi09,Ayi10,Yil11,Yil14b,Lac15}, we present here the first application to a realistic physical phenomena where the phase-space sampling is explicitly made.

We consider the SF of $^{258}$Fm that was used as a benchmark for TDDFT \cite{Sca15,Tan15}. The adiabatic energy landscape is shown in Fig. 1
of Ref. \cite{Tan15}
and is given with additional information in Fig. 1 of \cite{Sup17}.
This landscape was obtained using the HF+BCS approximation with the EV8 program \cite{Bon05}. 
The SLy4d interaction \cite{Kim97} is used in the mean-field together with a constant-$G$ pairing as in Refs. \cite{Sca15,Tan15}. 
Following a strategy similar to Ref. \cite{Sad16}, we suppose that the SF can be separated into two steps.  
First, the tunneling through the fission barrier up to a deformation $Q^{\rm ini}_2$ and, second,
the evolution through scission leading to fission. We apply our method to describe the second step.
We assume that the first step is sufficiently 
slow so that the adiabatic limit is meaningful. We further suppose that all the extra energy above the potential energy is converted into internal excitation energy $E^*$ of the fissioning nucleus. 
The excitation energies taken in our calculations are illustrated for different $Q^{\rm ini}_2$ in Fig. 1 of \cite{Sup17},  
assuming that the total energy is 1 MeV above the DFT ground state located at $Q^{\rm GS}_2=32$ barn. 

 For a given $Q^{\rm ini}_2$ value, the excitation induces internal fluctuations in single-particle DOFs. To mimic these fluctuations, we use the stochastic mean-field
technique \cite{Ayi08}. We consider the adiabatic quasi-particle vacuum $| \Psi (Q^{\rm ini}_2) \rangle$ and associated  one-body density 
$\rho=  \sum_{i} | \varphi_i \rangle n_i \langle \varphi_i |$.
$\{ | \varphi_i \rangle \}$ denotes the complete canonical basis.  From this information, one can construct an ensemble of one-body densities $\rho^{(n)}(t_0)$, where $(n)$ labels a specific event, followed by a set of independent TDDFT evolutions.  
The statistical properties of $\rho^{(n)}$ are given by following the original prescription \cite{Ayi08} 
and assuming that the one-body density is given by $\rho^{(n)} (t)  =  \sum_{ij} | \varphi_i \rangle \rho^{(n)}_{ij} \langle \varphi_j |$, where $\rho^{(n)}_{ij}$ are Gaussian random numbers verifying $\overline{\rho^{(n)}_{ij}} = \delta_{ij} n_i$ and 
\begin{eqnarray}
\overline{\delta\rho^{(n)}_{ij}\delta\rho^{(n)*}_{i'j'}}=\frac{1}{2}\delta_{ii'}\delta_{jj'}[n_i(1-n_j)+n_j(1-n_i)]. 
\label{eq:moment2}
\end{eqnarray}
 $\delta \rho^{(n)}$ denotes here the deviation around the mean value. Thus there are as many gaussian random numbers as the number of components $\rho_{ij}$ such that $n_i (1-n_j)$ or $n_j (1-n_i)$ is non-zero. 
We suppose in practice that fluctuations only occur between single-particle states in a narrow window of energy $\Delta \epsilon$ centered at the Fermi energy. The window size is fixed using energetic argument: for each initial condition, one calculates the associated DFT energy ${\cal E} (\rho^{(n)})$. The window is then adjusted so that the phase-space average on the energy fulfill the condition 
$E^*(Q^{\rm ini}_2) \simeq  \overline{{\cal E } (\rho^{(n)})} - {\cal E} (\overline{\rho^{(n)}})$.
Here, the last term identifies with the adiabatic energy. The energy windows, taken equal for proton and neutron, are displayed in Fig. 1 of \cite{Sup17}. 
Here, we have neglected possible fluctuations in the anomalous density \cite{Lac13}. 

In our approach, fluctuations only stems from the fluctuations in the initial density. Each initial density is then 
evolved in time using the TDDFT solver.
The evolutions have been performed including dynamical pairing using the TDHF3D+BCS code \cite{Sca13} generalized to treat non-diagonal matrix elements of the one-body density. 
The treatment of full density matrix requires extra numerical 
efforts due to the increase of complexity in the different fields entering in the functional. Generalized expressions of these fields are given in \cite{Sup17}. Note that the TDHF+BCS approach is obtained from TDHFB by neglecting some components of TDHFB. This approximation 
leads to specific difficulties \cite{Sca12}. It however includes in a reasonable way pairing effects on static nuclear properties and solves the threshold anomaly. By reducing significantly the numerical cost compared to TDHFB, it appears today as the best compromise to envisage several hundreds of trajectories as proposed here.
The coordinate space is discretized with a mesh size of $\Delta r=0.8$ fm within a box of the size $48.8\times48.8\times26.4$ fm$^3$. 
The time step in the dynamic is taken to be $\Delta t=1.5\times10^{-24}~{\rm s}$.

Initial fluctuations propagate in time and lead to a variety of final density profiles. 
Eq. (\ref{eq:moment2}) does not pre-suppose that selected DOFs contain more information than others.  
Any type of deformation can be accessed in time and most spacial symmetries can be spontaneously broken.  
\begin{figure}
\begin{center}
\includegraphics[scale=.6,angle= 0]{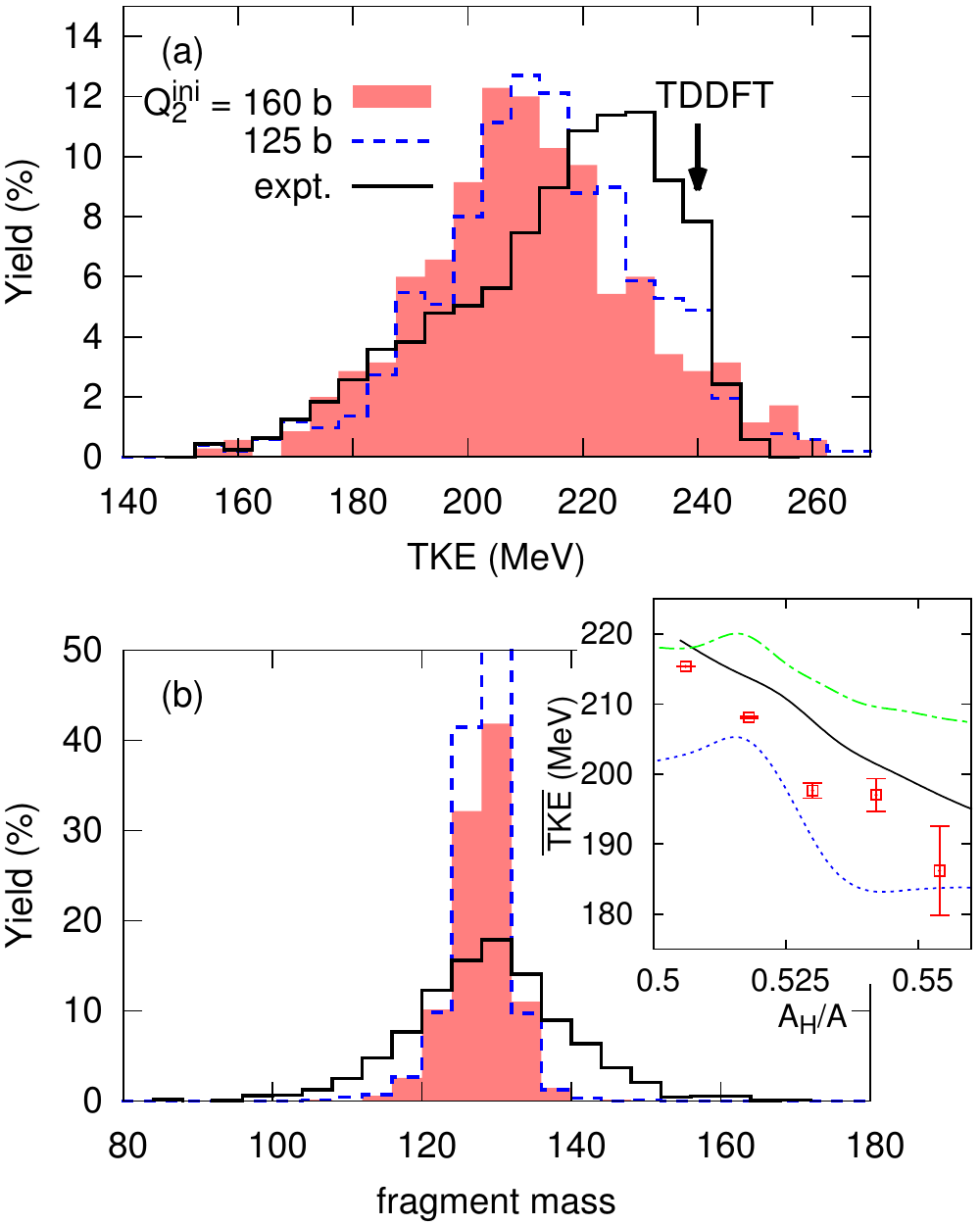}
\end{center}
\caption{ TKE  (a) and fragment mass (b) distributions obtained starting from 
$Q^{\rm ini}_2=160$ barn (shaded area) and 
$125$ barn (dashed line). The solid line is the experimental data \cite{Hul89}. In (a) the arrow indicated the mean TKE obtained in \cite{Sca15}. 
In the inset is shown the correlation between the average TKE and heaviest fragment mass (red squares). 
Comparison is made with results of the scission point model \cite{Wil76} in dotted and 
dot-dashed lines and $^{257}$Fm data \cite{Bal71} in solid line. }
\label{fig:tke}
\end{figure}
Besides these improvements, probability distributions of any 
one-body observables $A$, can be obtained 
using the set of values $A^{(n)} = {\rm Tr}(A \rho^{(n)})$. Mean values $\overline{A}$ and fluctuation $\sigma^2_A$ can then be also deduced by performing the classical phase-space average over different trajectories. 
350 and 512 TDDFT trajectories have been performed for $Q_2^{\rm ini}=160$ and 125 barn, respectively. The lowest value is inside the region where TDDFT without pairing does not lead to fission. At this position many single-particle crossing occurs (see Fig. 1 of Ref. \cite{Sup17}). The second deformation is at the TDDFT fission threshold and most single-particle crossing already occurred at lower deformation. Accordingly, we were anticipating very different final TKE and mass distribution depending on $Q_2^{\rm ini}$. This is not what 
we observed when quantum fluctuations are included.
We assume that the system has fissioned if the distance between the two fragments reaches $26$ fm before $t=4500$ fm/$c$. 

For each fissioning event, one can get the masses of fragments and by adding the Coulomb energy, reconstruct the TKE. 
The TKE and fragment mass distributions after the fission of $^{258}$Fm are compared in Fig. \ref{fig:tke} to the experimental data of Ref. \cite{Hul89}. The TKE distribution is well reproduced as well as its correlation with the heaviest fragment mass (inset of Fig. \ref{fig:tke}). For masses, while fluctuations are increased by a factor of 2 compared 
to the original TDDFT, the results still underestimate most asymmetric fission. It is interesting to observe that the TKE can be fairly well reproduced without invoking the role of the asymmetric fission mode as it is usually assumed \cite{Bon06}. Even if the reproduction of mass 
is a semi-success, to the best of our knowledge, this is the first time the TKE of SF is reproduced by a fully microscopic theory.  The mean and variance of TKE obtained are $\overline{E_{\rm TKE}}=211$ MeV and $\sigma_{E}=19$ MeV to be compared with the experimental references 
$\overline{E_{\rm TKE}}=215.5$ MeV and $\sigma_{E}=19.3$ MeV. We also display in Fig. \ref{fig:tke} the distributions obtained for $Q^{\rm ini}_2=125$ barn. This illustrates that the distribution is almost insensitive to the starting configuration over a rather large range of initial deformation. However, if the $Q_2^{\rm ini}$ is taken closer to the scission  point, as shown in Fig. 3 of \cite{Sup17}, 
the agreement of TKE distribution with the data deteriorates, indicating the subtle effect of dissipation leading 
to a complex balance between internal excitation and fragment acceleration before reaching scission.

It is worth mentioning that the effect of dynamical pairing is significantly washed out when including initial fluctuations. To illustrate this, we also performed evolutions with initial fluctuations but neglecting dynamical pairing, i.e. we froze the occupation numbers during evolution. This case, referred below as "without dynamical pairing", is systematically shown below. A first conclusion, is that, even if dynamical pairing is now neglected, the account for initial fluctuation is also enough to solve the threshold anomaly contrary to TDDFT without initial fluctuations. Indeed, by construction initial density fluctuations also induce jumps in single-particle space leading to fission. A second conclusion is that the fragment TKE and mass yields are essentially unaffected by dynamical pairing as soon as initial fluctuations are included (see Fig. 2 of \cite{Sup17}).
\begin{figure}
\begin{center}
\includegraphics[scale=.55,angle=0]{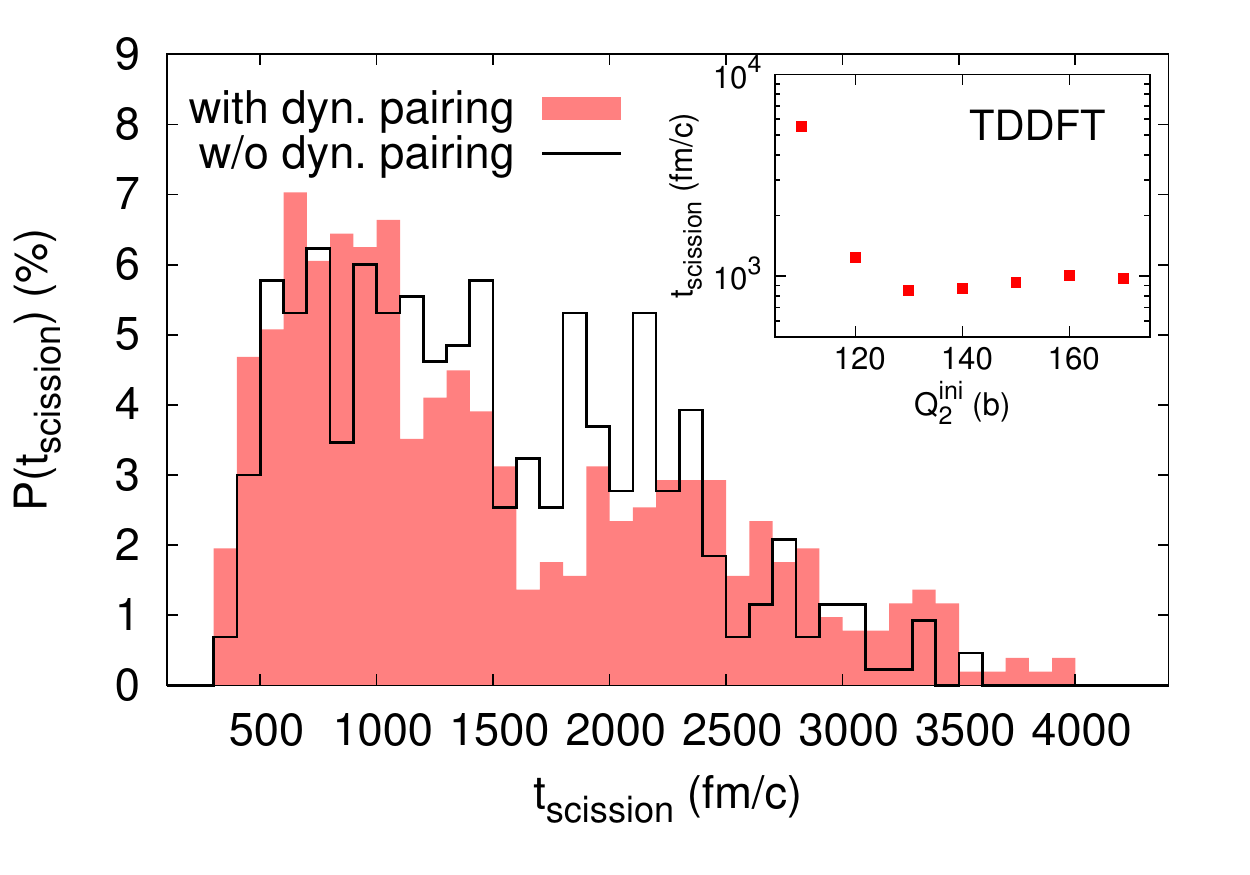}
\end{center}
\caption{Distribution of the time elapsed from $Q^{\rm ini}_2=125$ barn to the scission point including  (shaded area) or not (solid line) dynamical pairing. The time to reach scission using TDDFT with dynamical pairing and 
without initial fluctuation is shown in the inset as a function of initial deformation.}
\label{fig:time}
\end{figure}

\begin{figure}
\begin{center}
\includegraphics[scale=.7,angle=0]{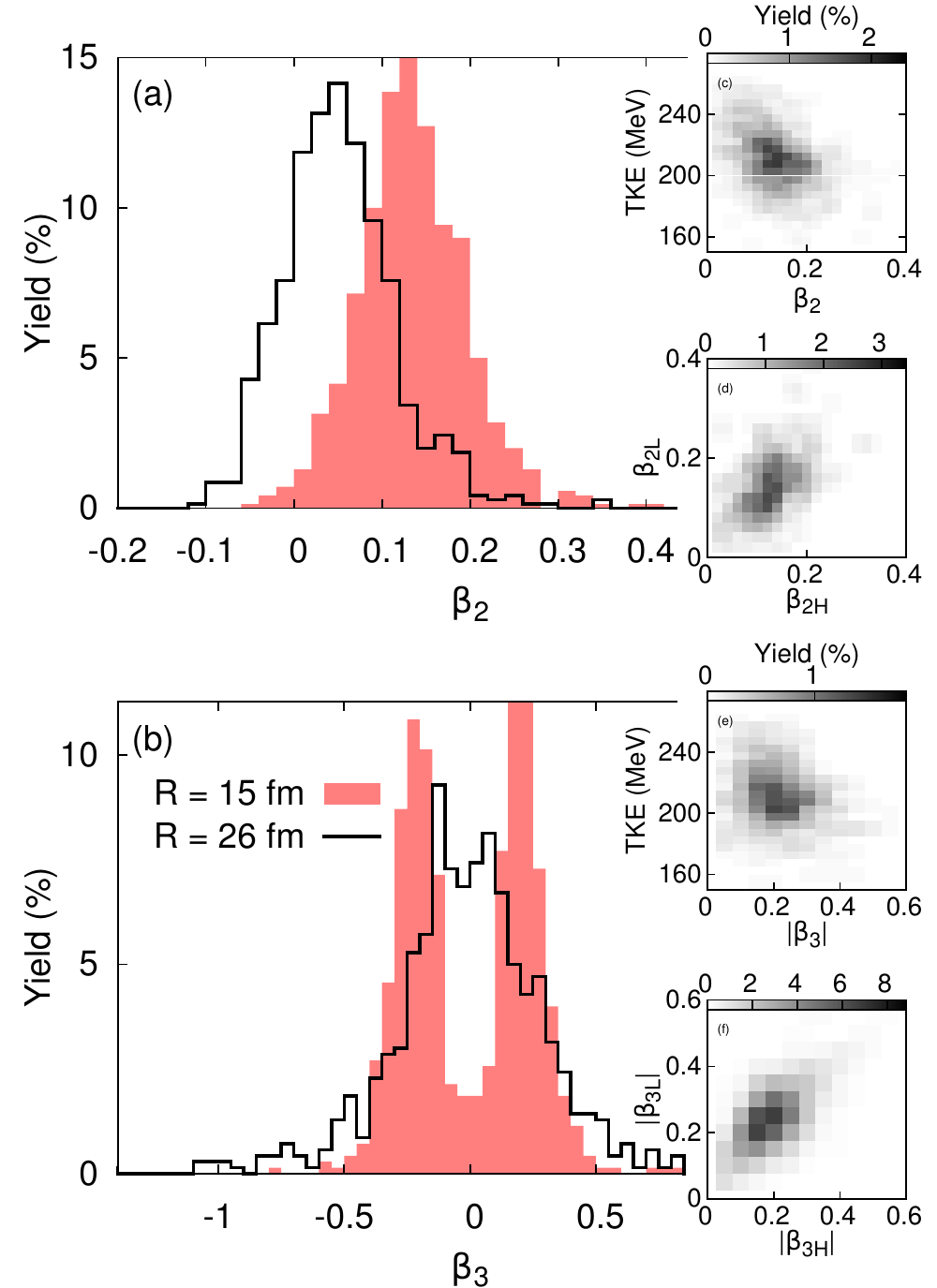} 
\end{center}
\caption{Calculated distributions of quadrupole $\beta_2$ (a) and octupole $\beta_3$ (b)  deformation parameters of fragments 
at scission point (shaded area) and at $26$ fm (black solid line). The insets (c) (resp. (e)) shows the correlation between 
the final TKE and $\beta_2$ (resp. $\beta_3$) at scission. The insets (d) (resp. (f)) shows the event by event correlation at scission 
between the $\beta_2$ (resp. $\beta_3$) of the heaviest (H) nucleus and the $\beta_2$ (resp. $\beta_3$) of the lightest (L) nucleus. }
\label{fig:q2q3-2}
\end{figure}

Important aspects that are scarcely known experimentally can be inferred from our calculation. 
We show in Fig. \ref{fig:time}, the distribution of time needed to reach scission point. We see 
two bumps in the time distribution that might stems from the non-trivial behavior of the scission time as a function of 
$Q^{\rm ini}_2$ already in TDDFT without fluctuations (inset of Fig. \ref{fig:time}). Indeed, after a sharp decrease of this time up to $Q^{\rm ini}_2\simeq 130$ b, 
this time increases again up to 160 b and then decreases again.
This underlines the complexity of the collective paths that stems from the possibility to access various shapes during the evolution leading to energy exchange between collective and single-particle DOFs and ultimately to dissipation. 
To illustrate this effect, we display in Fig. \ref{fig:q2q3-2}, the distribution of quadrupole $\beta_2$ and octupole $\beta_3$ deformation parameters  \cite{Tak98,Yam01} of fragments. 
Fragments are deformed at and close to the scission point. This deformation relaxes as the two fragments escapes from each others. 
Several interesting features are seen in Fig. \ref{fig:q2q3-2}. 
First, the scission preferentially occurs when both fragments are prolate together with large octupole deformation.  
In addition, from the insets of Fig. \ref{fig:q2q3-2}, a strong correlation between the final TKE and the deformations of fragments is seen. 
The deformation behaviors directly indicate that a part of TKE dissipates into excitation energies of the fragments, 
which makes the large variation of the final TKE measured experimentally. 

\begin{figure}
\begin{center}
\includegraphics[scale=.3,angle=-90]{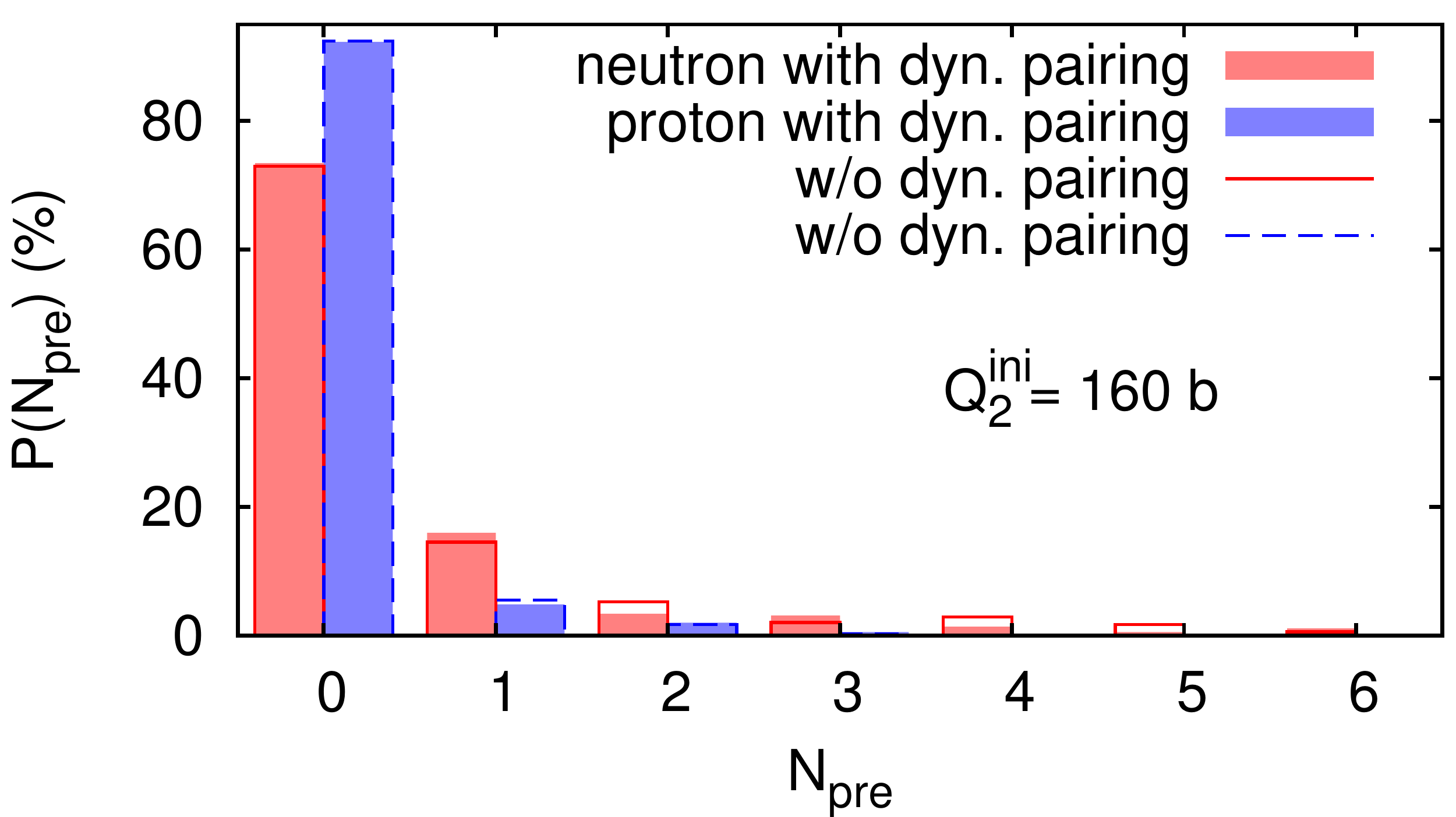}
\end{center}
\caption{Probability distribution of the number $N_{\rm pre}$ of neutrons (red shaded area) and protons (blue shaded area) 
emitted before  reaching scission. Neutron (resp. proton) distribution are shifted by -0.5 (resp +0.5) unit for display reasons.   
The equivalent distribution obtained without dynamical pairing are also shown for neutron (solid line)
and proton (dashed line).}
\label{fig:evapnp}
\end{figure}

Since the system is excited, it can cool down by particle emission. In TDDFT, particles can be emitted to the continuum. These particles are then removed from the calculation by adding a small absorbing imaginary potential at the boundary of the mesh leading to a decreasing of the total mass $A^{(n)}(t) = {\rm Tr}(\rho^{(n)}(t))$. The number of evaporated particles  as a function of time is then estimated event-by event by simply making the difference between 
$A^{(n)}(t) $ and the initial mass.
Fig. \ref{fig:evapnp} shows the probability distribution of the number of emitted
protons and neutrons before scission. Note that this procedure leads to a continuous distribution of mass, with eventually non-integer values of $A^{(n)}$ at final time. The distribution in Fig. \ref{fig:evapnp} is obtained assuming a binning $\Delta A$ of one unit mass around integer mass values.
Neutron emission is obviously favored due to (i) the absence of Coulomb barrier, (ii) the favorable $N/Z$ ratio of the fissioning 
nucleus.   We see that there is a non negligible chance to emit particles in the early stage of fission.  

In this letter, we present a novel microscopic approach to SF.  We assume that, after the system passes the barrier, the initial 
phase-space explored in collective space is fixed by simple hypothesis on energy conservation.  
This approach, applied to $^{258}$Fm  
is providing for the first time a fully microscopic description of the fragment 
TKE distribution after fission and gives unique microscopic information on the fission process.  Scission time fluctuations, intrinsic deformation, pre-scission neutron/proton emissions are analyzed. We note that the mass asymmetry after fission is still underestimated. This could be traced back to the energy criteria used to sample initial conditions. Indeed, from Fig. 2 of Ref. \cite{Sta09}, we see that the initial energy is too low to classically access the asymmetric path. Increasing the energy should allow more asymmetric shape. To check this hypothesis we also performed a set of TDDFT calculation with $Q^{\rm ini}_2=160$ b and higher 
average excitation energy (see Fig. 4 of \cite{Sup17}). The mass asymmetry is much better reproduced while the TKE is shifted to lower energy. This result is very promising although it also shows that getting back more asymmetric fission might degrade the agreement on TKE. 
Increasing the energy is however not justified and one should normally include the proper quantum weight of different paths. For SF, this weight is directly linked to the tunneling probability through the collective fission barrier. In the near future, it might be interesting to couple our approach with the method used in Ref. \cite{Sad16} to obtain the tunneling probability. 
Another important issue to be clarified is the dependence of the results with respect to the parameters entering in the functional and for instance to redo the same calculation using for instance globally optimized UNEDFx functionals \cite{Kor10,Kor14}. 

We thank B. Yilmaz for continuous discussions during this work and D. Regnier for discussions on TDGM.
S.A. gratefully acknowledges the IPN-Orsay for the partial support and the warm hospitality extended to him during his visits. 
This work is supported in part by the US DOE Grant No. DE-SC0015513. This project has received funding from
the European Unions Horizon 2020 research and innovation
program under grant agreement No. 654002.

\end{document}